\documentclass[preprint]{elsarticle}
\usepackage{lineno,hyperref,amssymb}
\pdfoutput=1

\makeatletter
\def\@xfootnote[#1]{%
  \protected@xdef\@thefnmark{#1}%
  \@footnotemark\@footnotetext}
\makeatother

\journal{Nuclear Instrumentation and Methods B}

\begin{document}

\begin{frontmatter}

\title{Improvements to TITAN's Mass Measurement and Decay Spectroscopy Capabilities}

\author[TRIUMF]{D. Lascar\corref{correspondingauthor}}
\cortext[correspondingauthor]{Corresponding author. Tel: +1-604-222-1047 x6815; Fax: +1-604-222-1074}
\ead{dlascar@triumf.ca}

\author[TRIUMF,TAMU]{A.A. Kwiatkowski}
\author[Munster]{M. Alanssari}
\author[TRIUMF,Manitoba]{U. Chowdhury}
\author[TRIUMF]{J. Even}
\author[TRIUMF,UBC]{A. Finlay}
\author[TRIUMF,UBC]{A.T. Gallant}
\author[TRIUMF]{M. Good}
\author[TRIUMF,MPI]{R. Klawitter}
\author[TRIUMF,Manitoba]{B. Kootte}
\author[TRIUMF,Water]{T.Li}
\author[TRIUMF,SFU,Mines]{K.G. Leach}
\author[TRIUMF,Munster]{A. Lennarz}
\author[TRIUMF,UBC]{E. Leistenschneider}
\author[Calgary]{A. J. Mayer}
\author[ND]{B.E. Schultz}
\author[TRIUMF,MPI]{R. Schupp}
\author[TRIUMF,SFU]{D.A. Short}
\author[SFU]{C. Andreoiu}
\author[TRIUMF,UBC]{J. Dilling}
\author[Manitoba]{G. Gwinner}

\address[TRIUMF]{TRIUMF, 4004 Wesbrook Mall, Vancouver, British Columbia V6T 2A3, Canada}
\address[TAMU]{Cyclotron Institute, Texas A\&M University, College Station, Texas 77843, USA}
\address[Munster]{Institut f\"{u}r Kernphysik, Westf\"{a}lische Wilhelms-Universit\"{a}t, 48149 M\"{u}nster, Germany}
\address[Manitoba]{Dept. of Physics \& Astronomy, University of Manitoba, Winnipeg, Manitoba R3T 2N2, Canada}
\address[UBC]{Dept. of Physics \& Astronomy, University of British Columbia, Vancouver, British Columbia V6T 1Z1, Canada}
\address[MPI]{Max-Planck-Institut f\"{u}r Kernphysik, Heidelberg D-69117, Germany}
\address[Water]{Dept of Physics \& Astronomy, University of Waterloo, Waterloo, Ontario  N2L 3G1, Canada}
\address[SFU]{Dept. of Chemistry, Simon Fraser University, Burnaby, British Columbia V5A 1S6, Canada}
\address[Mines]{Dept. of Physics, Colorado School of Mines, Golden, Colorado, 80401, USA}
\address[Calgary]{Dept. of Physics \& Astronomy, University of Calgary, Calgary, Alberta T2N 1N4, Canada}
\address[ND]{Dept. of Physics, University of Notre Dame, Notre Dame, Indiana 46556, USA}


\begin{abstract}
The study of nuclei farther from the valley of $\beta$-stability than ever before goes hand-in-hand with shorter-lived nuclei produced in smaller abundances than their less exotic counterparts. The measurement, to high precision, of nuclear masses therefore requires innovations in technique in order to keep up. TRIUMF's Ion Trap for Atomic and Nuclear science (TITAN) facility deploys three ion traps, with a fourth in the commissioning phase, to perform and support Penning trap mass spectrometry and in-trap decay spectroscopy on some of the shortest-lived nuclei ever studied. We report on recent advances and updates to the TITAN facility since the 2012 EMIS Conference.

TITAN's charge breeding capabilities have been improved and in-trap decay spectroscopy can be performed in TITAN's Electron Beam Ion Trap (EBIT). Higher charge states can improve the precision of mass measurements, reduce the beam-time requirements for a given measurement, improve beam purity and open the door to access isotopes not available from the ISOL method via in-trap decay and recapture. This was recently demonstrated during TITAN's mass measurement of $^{30}$Al. The EBIT's decay spectroscopy setup was commissioned with a successful branching ratio and half-life measurement of $^{124}$Cs. Charge breeding in the EBIT increases the energy spread of the ion bunch sent to the Penning trap for mass measurement, so a new Cooler Penning Trap (CPET), which aims to cool highly charged ions with an electron plasma, is undergoing offline commissioning. Already CPET has demonstrated the trapping and self-cooling of a room-temperature electron plasma that was stored for several minutes. A new detector has been installed inside the CPET magnetic field which will allow for in-magnet charged particle detection.
\end{abstract}

\begin{keyword}
Mass Spectrometry\sep Ion Trapping\sep Ion Cooling\sep HCI
\end{keyword}

\end{frontmatter}

Precision mass measurements provide critical input data for nuclear astrophysics models \cite{Schatz2013}, nuclear structure calculations \cite{Lunney2003}, and tests of fundamental symmetries \cite{Diffusion1990}. The precision requirements for each field varies from $\frac{\delta m}{m} \leq 10^{-7}$ for nuclear astrophysics to $\frac{\delta m}{m} \leq 10^{-10}$ for tests of Charge, Parity and Time (CPT) invariance. Penning traps are the tool of choice for online precision mass determinations \cite{Blaum2012} and have demonstrated the ability to quickly and precisely measure masses at a variety of different facilities with differing means of production \cite{Valverde2015,VanSchelt2013,Kankainen2013,Macdonald2014,Droese2013}. As newer facilities as well as upgrades to older facilities are coming online and providing access to nuclei further from the valley of $\beta$-stability than ever before, the need to make precise mass measurements using shorter measurement cycles than before is ever present.

TRIUMF's Ion Trap for Atomic and Nuclear science (TITAN) facility \cite{Dilling2003}, coupled to TRIUMF's Isotope Separator and ACcelerator (ISAC) in Vancouver, Canada, is one such facility capable of performing high precision mass measurements. At TITAN, three ion traps are used to perform and support high precision mass measurements or perform in-trap decay spectroscopy. Since the last EMIS conference in 2012 \cite{Kwiatkowski2013}, TITAN has made advances in the service of high-precision mass measurements and in-trap decay spectroscopy.

\section{The TITAN System}

Ions produced at ISAC are selected by a magnetic dipole mass separator ($R \approx 2500$ \cite{Dilling2014}) and delivered to the TITAN facility at a typical energy of 20 keV. The ions are thermalized and bunched in TITAN's He-filled radiofrequency quadrupole (RFQ) cooler-buncher \cite{Brunner2012a}.

After ejection the cooled ion bunch passes through an electrostatic beam switchyard where the ions can be sent either to TITAN's Electron Beam Ion Trap (EBIT) or directly into TITAN's Measurement Penning Trap (MPET) for precision mass measurement. Before entering MPET, all ions pass through TITAN's Bradbury Nielsen Gate (BNG), which performs a mass selective deflection with $R\sim 100$ based on the ions' times of flight \cite{Brunner2012}. In the EBIT, ions can be charge-bred and then sent to MPET for mass measurement (see Section \ref{subsec:mass}), or remain in the EBIT to perform in-trap decay spectroscopy (see Section \ref{subsec:EBITspec}).

Precision mass measurements are carried out in MPET \cite{Brodeur2009,Brodeur2012} where a 3.7 T magnetic field confines ions radially and a harmonic electrostatic potential confines ions axially. Masses, $m$, are measured using the Time-of-Flight Ion-Cyclotron-Resonance (ToF-ICR) method which measures an ion's cyclotron frequency,
\begin{equation}
\omega_c = \frac{qeB}{m},
\end{equation}
where $B$ is the magnetic field in the Penning trap, $q$ is the ion's charge state and $e$ is the elemental charge of $1.6\times 10^{-19}$ C. Through an optimization of both ion injection optics into MPET as well as an efficient measurement cycle using a Lorentz-steerer technique \cite{Ringle2007}, TITAN has specialized in measuring the masses of the shortest-lived nuclides ever measured. Among the shortest-lived species successfully measured are $^{31}$Na ($t_{1/2}$ = 17 ms) \cite{Chaudhuri2013}, $^{32}$Na ($t_{1/2}$ = 13.2 ms) [\emph{paper in process}], and $^{11}$Li ($t_{1/2}$ = 8.75 ms) \cite{Smith2008}. A measurement cycle time of 6.7 ms with a 6 ms quadrupolar excitation time has been shown to be the shortest measurement cycle time so far \cite{Chaudhuri2014}.

\subsection{Improving Mass Measurement Precision}

A typical Penning trap mass measurement has a fractional statistical precision,
\begin{equation}
\frac{\delta m}{m} \propto \frac{m}{qeBt_{\mathrm{rf}}\sqrt{N_{\mathrm{ion}}}},
\label{eq:precision}
\end{equation}
where $m$ is the mass of the ion of interest and $t_{\mathrm{rf}}$ is the quadrupole excitation time for the ion in the Penning trap. For more stable isotopes, increasing the quadrupole excitation time is a trivial means of improving the precision, but for short-lived isotopes that method is limited since the ion of interest may decay in the trap. In practice, the maximum applicable $t_{\mathrm{rf}} \approx 2t_{1/2}$ so for short-lived isotopes improvements in precision must come from other parameters.

The precision in Equation \ref{eq:precision} improves as the charge state increases with an obvious limit of $q=Z$. For example, an ion in a +2 charge state will resonate in the Penning trap at a frequency twice that of the singly charged ion (SCI) but the uncertainty of the frequency measurement, $\delta \omega$, is (all things being equal) unchanged. Since $\frac{\delta \omega}{\omega}=\frac{\delta m}{m}$, the precision of the mass measurement should improve by a factor of 2 by measuring a frequency that is twice as large.

\section{The TITAN EBIT}
\label{sec:EBIT}

TITAN's EBIT \cite{Lapierre2010} was designed to charge-breed ions in preparation for their mass measurement in MPET. Since the first HCI mass measurement of $^{44}$K$^{4+}$ \cite{Lapierre2010} and the first short-lived HCI mass measurement of $^{74}$Rb$^{8+}$ \cite{Ettenauer2011}, TITAN has measured and published the masses of 22 exotic ground-state nuclei \cite{Macdonald2014,Ettenauer2011,Frekers2013,Lapierre2012,Gallant2012a,Simon2012} and the long-lived isomeric state of $^{78}$Rb \cite{Gallant2012a}. The EBIT consists of a superconducting magnet in a Helmholtz configuration capable of generating a magnetic field of up to 6 T. An electron gun typically produces electron beam energies of 1.5-7 keV and currents as high as 500 mA\footnote{Though the gun was designed to provide a 70 keV beam at 5 A.}. The EBIT is surrounded by seven radial ports with recessed Beryllium windows that can accommodate either Lithium-drifted Silicon (Si(Li)) or high-purity Germanium (HPGe) detectors \cite{Leach2015}. With these detectors in place, in-trap decay spectroscopy and half-life measurements such as those of $^{124}$Cs and $^{124}$In \cite{Lennarz2014} have been performed. The ability to capture multiple ion bunches inside the EBIT up to the trap's space charge limit of $10^9 e$ was also demonstrated during the $^{124}$Cs and $^{124}$In measurements \cite{Leach2015a,Klawitter2015a}.

When charge-breeding for mass measurements in MPET, the EBIT's electron beam imparts an estimated energy spread of 10-100 eV/charge on the highly charged ion (HCI) bunch \cite{Ke2007}. An increased energy spread increases the emittance of the ion cloud, adversely affects the trapping potential in MPET, decreases the trapping efficiency in MPET and has the overall effect of reducing the precision of mass measurements. We are currently commissioning a new Cooler PEnning Trap (CPET) \cite{Ke2007} whose purpose is to cool HCI bunches to $\sim$1 eV/charge using a plasma of trapped, self-cooled electrons while avoiding charge exchange with the cooling medium (see Section \ref{sec:CPET}).

\subsection{EBIT for mass measurement support}
\label{subsec:mass}

\begin{figure}[htb]
    \begin{center}
        a)
        \includegraphics[width=7cm]{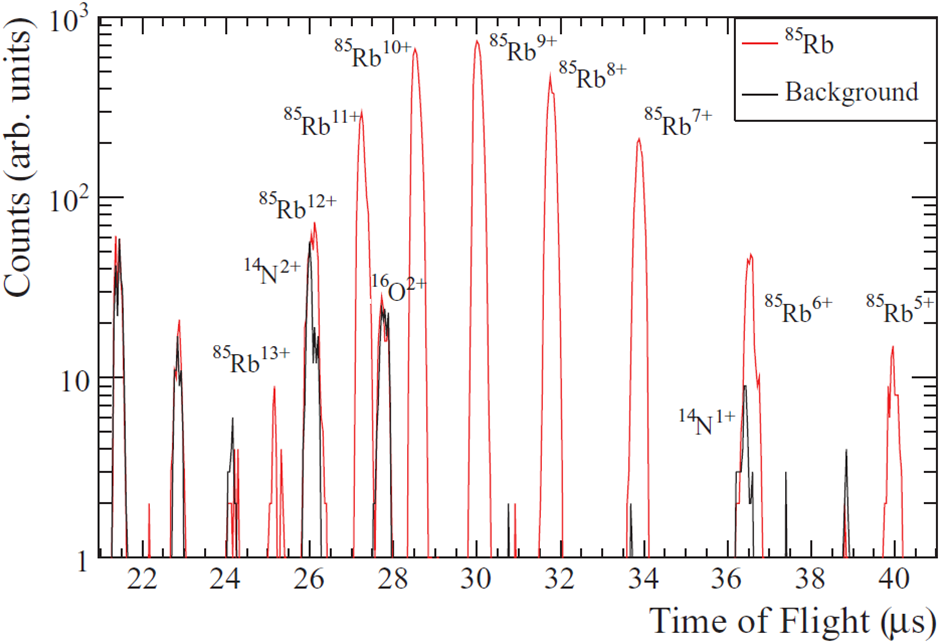}\\
        b)
        \includegraphics[width=7cm]{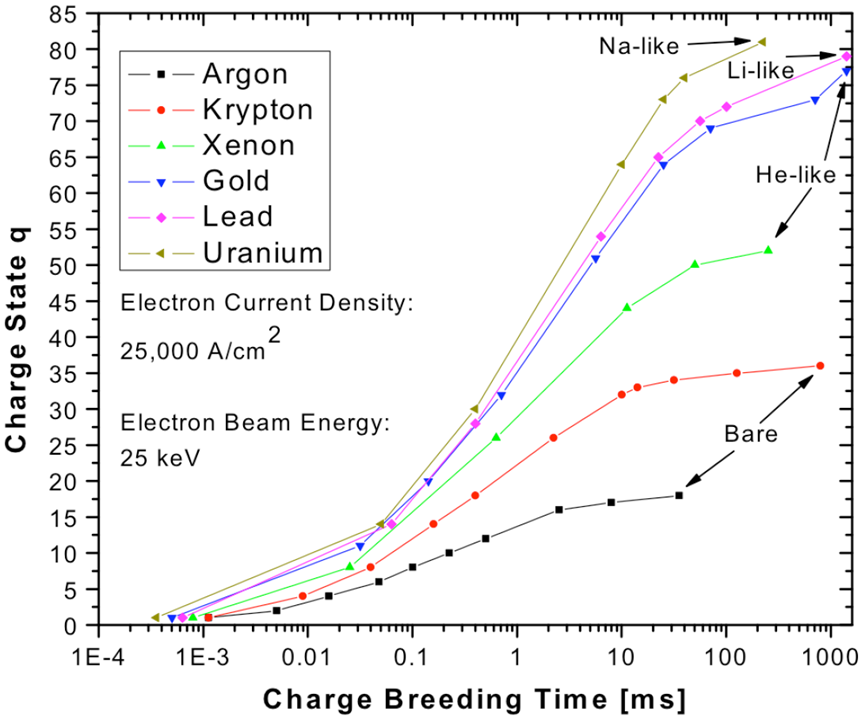}
        \caption[ToF separation of charge bred ions]{a) $t_{\mathrm{cb}} = 23$ ms. ToF spectra of highly charged ions extracted from the EBIT with and without injected Rb ions \cite{Gallant2012a}. b) Calculated charge breeding times of high Z elements \cite{Froese2006a}.  (Color online)}
        \label{fig:separation}
    \end{center}
\end{figure}

The EBIT has been used successfully to enhance the mass measurements of, for example, $^{74}$Rb where the higher charge state of $8^+$ was used to improve the mass measurement precision \cite{Ettenauer2011} or $^{71}$Ge and $^{71}$Ga where charge breeding was used to aid in mass selection \cite{Frekers2013}. While Equation \ref{eq:precision} demonstrates the statistical improvement in precision from using HCI $\left(\delta m/m \propto q^{-1}\right)$, that relation only dominates as long as ion production is high enough that 1-5 ions can be sent into MPET for a given measurement cycle. If less than an ion/shot is produced the statistical gains from working with HCI can be outweighed by the $\sqrt{N_{\mathrm{ion}}}$ dependence in Equation \ref{eq:precision}. The effective gain in precision from using HCI, $G_{\mathrm{HCI}}$, is therefore given by,
\begin{equation}
G_{\mathrm{HCI}} = q \sqrt{2^{-t_{\mathrm{cb}}/t_{1/2}}\eta_{\mathrm{pop}}\epsilon},
\label{eq:gain}
\end{equation}
where $t_{\mathrm{cb}}/t_{1/2}$ is the ratio of the charge-breeding time to the nuclear half-life of the isotope undergoing charge-breeding in the EBIT, $\eta_{\mathrm{pop}}$ is the fraction of the ion's population in the bunch with charge state $q$, and $\epsilon$ is the combined efficiencies related to the use of HCI including (but not limited to) transport, capture in EBIT, capture in MPET, losses in EBIT due to charge-breeding and loss of charge state due to charge exchange with residual gas in both the EBIT and MPET \cite{Ettenauer2013}. $\epsilon$ is experimentally measured and, with the current state of the system, is on the order of 1\%. To charge-breed ions to a charge state of 10-20$^+$ requires between 0.01ms and 10 ms depending on the $Z$ of the element being charge-bred for $Z\geq30$ under typical EBIT operation settings \cite{Froese2006a}. $t_{\mathrm{cb}}$ to 10$^+$-20$^+$ is inversely proportional to $Z$ (see Figure \ref{fig:separation}b). The time-separation of the charge bred ions at MPET after ejection from the EBIT is $2-3$ $\mu$s for $A/q = 5-10$ (see Figure \ref{fig:separation}a).

Another advantage of using HCI is that mass measurements can be made in less time because the fractional precision of the measurement is also related to $\sqrt{N_{\mathrm{ion}}}$. For example, if a measurement with SCI would normally take 8 hours then the same measurement with the same $t_{\mathrm{rf}}$ would take 4 hours at a charge state of $4^+$ and 2 hours at a charge state of $16^+$ as long as $t_{1/2}\gg t_{\mathrm{cb}}$. Furthermore, the shorter duration measurement with the $16^+$ charge state will yield a result of the same precision as the original longer duration measurement with the $1^+$ charge state. This reduces the beam-time requirements for making precision measurements and makes experiments more attractive to a facility's experimental advisory committee.

\subsection{In-Trap Decay Spectroscopy in the EBIT}
\label{subsec:EBITspec}
\begin{figure}[htb]
    \begin{center}
        \includegraphics[width=5cm]{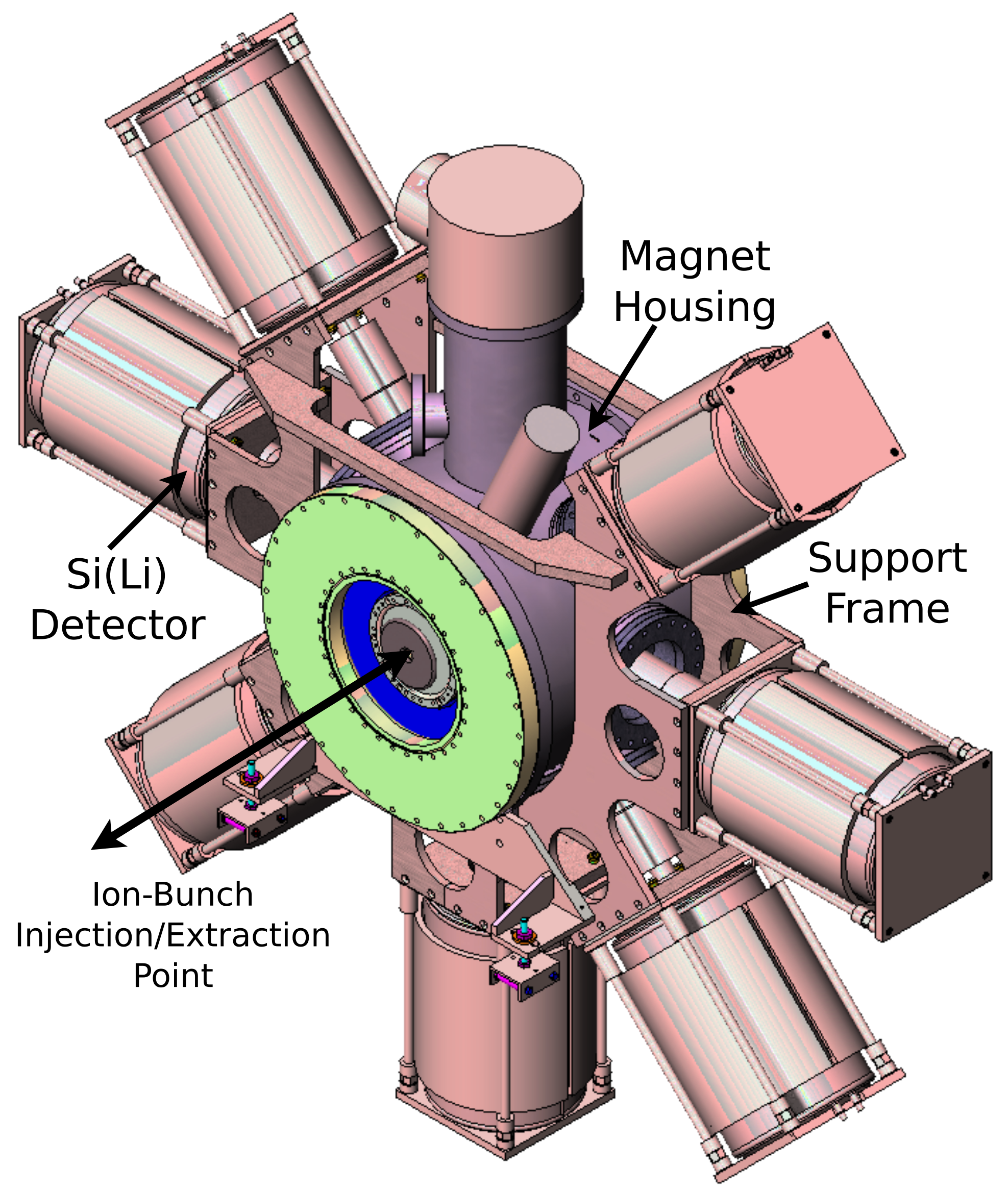}
        \caption[EBIT detector layout]{The TITAN EBIT in a seven-Si(Li) detector configuration. The top port is occupied by the EBIT's cryocooler. (Color online)}
        \label{fig:detectors}
    \end{center}
\end{figure}
The observation of neutrinoless double beta-decay ($0\nu\beta\beta$) would violate lepton number conservation \cite{Valle1982,Takasugi1984}, requiring the neutrino to be its own anti-particle, like a Majorana-particle. The effective Majorana-mass, $\langle m_{\beta\beta}\rangle$, of the neutrino could be extracted from:
\begin{equation}
\left(t^{0\nu\beta\beta}_{1/2}\right)^{-1} = G^{0\nu\beta\beta} (Q,Z)|M^{0\nu\beta\beta}|^{2}\langle m_{\beta\beta}\rangle^2
\label{eq:decayhalf}
\end{equation}
where $t^{0\nu\beta\beta}_{1/2}$ is the observed half-life of the $0\nu\beta\beta$ decay and $G^{0\nu\beta\beta} (Q,Z)$ is the phase space factor. $M^{0\nu\beta\beta}$ is the theoretically calculated Nuclear Matrix Element (NME) connecting the initial and final states \cite{Barea2012}. Any constraints placed on the NME must come from experimental data including ``measurements of the $\beta^-$ and electron-capture (EC) branching ratios of intermediate nuclei in [the] $2\nu\beta\beta$-decay process'' \cite{Leach2015a}. Since EC is typically a weaker transition, relative to the dominant $\beta^-$, background suppression is key to measuring the EC half-life and branching ratio.

The energy of the characteristic photon emitted during EC relevant to $\beta\beta$-decay studies is typically $< 40$ keV. The photon is generated from the K-shell electronic vacancy and subsequent electronic collapse in the wake of an EC. At energies of 40 keV or below, photons from electron-positron annihilation and Compton scattering contribute to a large low-energy background that can only be reduced if the decay environment is tightly controlled; an environment which can be provided by ion traps.

The TITAN EBIT is an open-access ion trap with seven ports separated by $45^\circ$ from one another and each separated from atmosphere by thin (0.25 mm thickness) Be windows (see Figure \ref{fig:detectors}). The open-access design is suitable for coupling photon detectors to the trap; thus, allowing for decay spectroscopy experiments. Si(Li) detectors were chosen for their good energy resolution and detection efficiency for photons with energy $\leq 50$ keV \cite{Lapierre2010}.

With the Si(Li) detectors in place, the EBIT can be used to perform in-trap decay spectroscopy. The strong magnetic field eliminates the background signal of 511 keV photons from $\beta$-particle annihilation by guiding charged decay products--such as electrons and positrons--away from the focal plane of the Si(Li) detectors. The elimination of the 511 keV background is particularly important for measurements which require a high-level of sensitivity at low photon energies. The proof of principle for this method was shown with the decay half-life and branching ratio measurements of $^{124}$Cs and $^{124,m}$In into $^{124,m}$Xe and $^{124,m}$Sn, respectively \cite{Lennarz2014}.

Target chemistry plays a role in many of the decisions made when running an experiment at an ISOL facility. If a given element doesn't diffuse out of the target material quickly enough (or at all) due to the chemical processes involved then that element cannot be studied. If, however, a desired isotope is the daughter of a parent whose element can be extracted from an ISOL target then, as long as the parent can be captured and the daughter subsequently recaptured, a study can be undertaken. Furthermore, if the decay-daughter in question is an isomeric state, then in-target production is often even more difficult. The same recapture of the daughter could grant access to isomeric states previously unavailable at ISOL facilities. Into this regime enters the EBIT.

TITAN has demonstrated the principle of in-trap recapture in the EBIT with the capture of $^{30}$Mg ($t_{1/2} = 335$ ms) and the recapture of its daughter, $^{30}$Al ($t_{1/2} = 3.6$ s). $^{30}$Al was subsequently delivered to and uniquely identified in MPET for diagnostic purposes.

Most of the recapture in the EBIT is performed purely by the space charge effects from the electron beam and the remaining daughters are recaptured with the aid of the magnetic field. A sufficiently pure daughter beam from the EBIT can be produced simply by waiting for the appropriate number of parent half-lives. Charge breeding can occur in parallel with the recapture and, with the concurrent use of the threshold charge breeding technique, isobaric separation can also be accomplished. A paper by Kwiatkowski \emph{el al.} to be submitted to \emph{Physical Review C} is in process.

\section{Advances with TITAN's Cooler Penning Trap}
\label{sec:CPET}

One of the consequences of charge-breeding ions in the EBIT is the energy spread increase of the charge-bred bunch to 10-100 eV/charge. The increased energy spread decreases trapping efficiency in MPET and also broadens the ToF resonance in a ToF-ICR measurement, thereby decreasing the precision of mass measurements and limiting the gains that using HCIs should bring. One standard method of cooling ions is to use a buffer gas in an RFQ or Penning trap \cite{Savard1991}, but any ion in a charge state greater than $1^+$ or $2^+$ will undergo charge exchange with the cooling medium. TITAN is in the process of commissioning CPET, whose goal is to cool HCIs sympathetically via the Coulomb interaction using a trapped electron plasma \cite{Schultz2013,Simon2013}.

Radial confinement in the Penning trap is achieved via the use of a 7 T magnetic field. Electrons are advantageous in this cooling scheme because their small mass allows them to quickly self-cool in the magnetic field via the emission of synchrotron radiation in $\sim 100$ ms or less. When CPET is brought online its use and cooling time will have to be tuned based on the rates of production of the isotope under study and the isotope's lifetime.

%
%
\subsection{Initial CPET Results}

CPET has demonstrated the trapping and self-cooling of the non-neutral plasma in the trap. On its own, the plasma will cool to its minimum value in 2-3 seconds (see Figure 3 in \cite{Chowdhury2015}) and methods exist to reduce that cooling time via the application of nonresonant \emph{rf} to the quadrupole structures inside CPET \cite{Maero2011}.

With a phosphor screen inserted inside the trap structure, electron plasma trapping times of longer than 2 minutes have been observed. Further, it has been determined that the number of electrons captured and cooled in the trap is between $10^8 - 10^{10}$ electrons, reaching the cooling requirements calculated by Ke \emph{et al} \cite{Ke2007}.


\subsection{CPET's New Mesh Detector}

\begin{figure}[!h]
    \begin{center}
        a)
        \includegraphics[width=6cm]{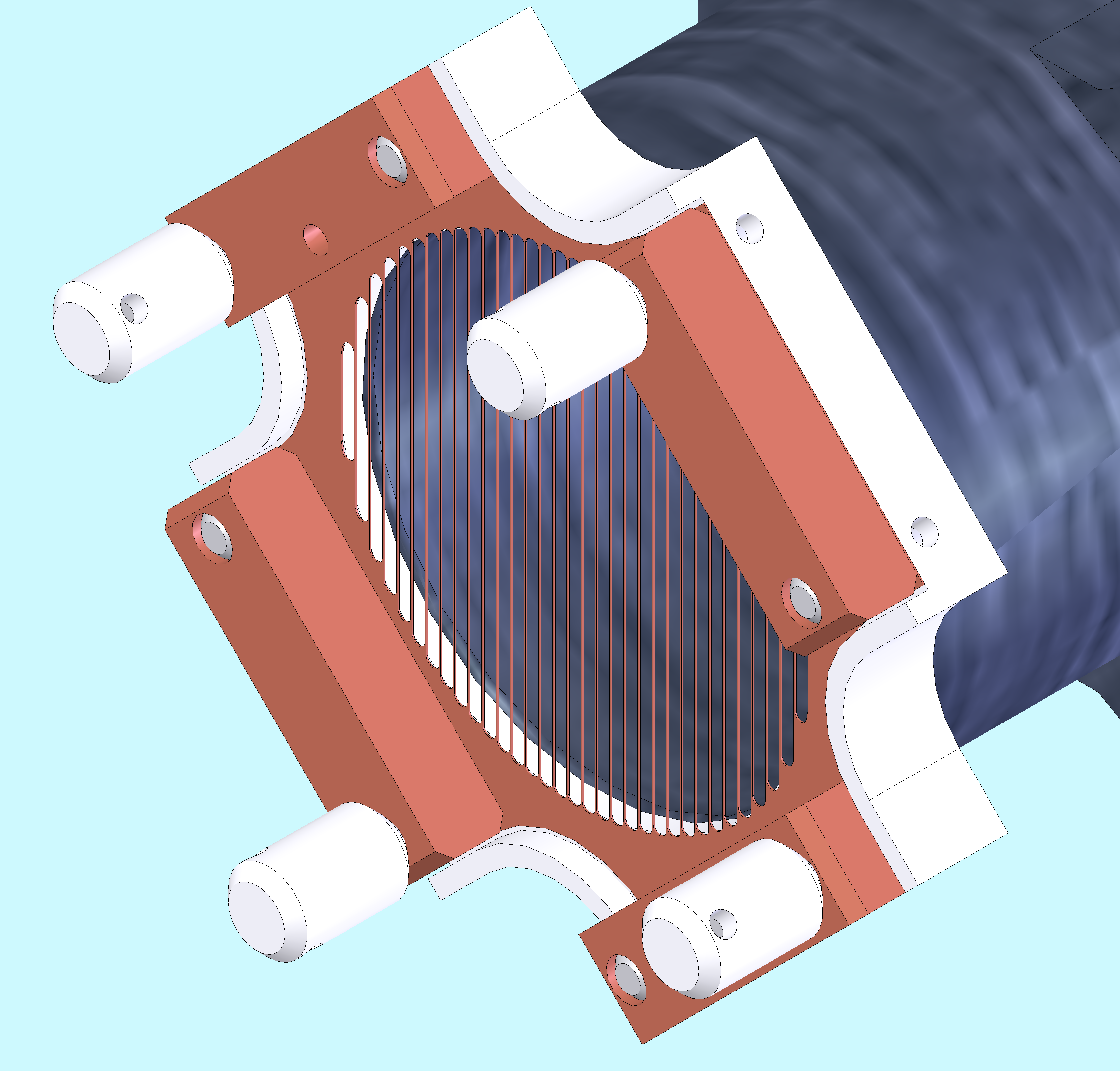}\\
        b)
        \includegraphics[width=8cm]{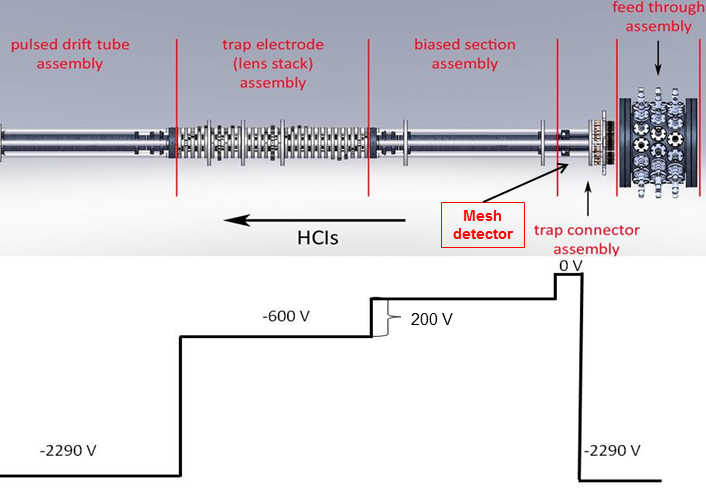}
        \caption[Mesh Detector]{a) The wires of CPET's mesh detector are laid out on a circle with a 34 mm diameter b) (\emph{Above}) - For reference, the ``trap electrode assembly'' is 400 mm in length. (\emph{Below}) - The trap's biasing scheme for when the mesh detector is used. (Color online)}
        \label{fig:mesh}
    \end{center}
\end{figure}

In order to move forward with the simultaneous trapping of ions and electrons, a new detection scheme for trapped electrons was required. The phosphor screen used to make the measurements in \cite{Chowdhury2015} was placed inside the CPET magnet to maximize detection efficiency and avoid the divergence of the electrons as they follow the field lines of the CPET magnet outside the trap. However, this effectively blocked the flight path for ions. A new detector was required for diagnostics that was capable of operating in the strong magnetic field of CPET. The diameter of the magnet's bore tube, which is less than 80 mm, made a removable detector impractical; hence a detector that was transparent from an electrostatic standpoint was required.

The detector design is based on the TITAN BNG \cite{Brunner2012}. An anode was made out of a 0.1 mm-thick copper plate with material photochemically etched away such that a unidirectional array of parallel, 0.1 mm-diameter wires remained (see Figure \ref{fig:mesh}a). The wires extend across a gap with a diameter of 34 mm and are separated by a 1 mm pitch. The thickness and pitch of the detector wires are identical to the dimensions of TITAN's BNG. With insulating ceramic holders, the entire detector assembly is just over an inch thick making its profile small and particularly easy to accommodate. This so-called ``mesh detector'' was installed at the end of CPET's entrance drift tube (See Figure \ref{fig:mesh}b).

When the detector is set to the experiment's electrical drift-tube potential, it is effectively transparent to charged particles passing through, and only the physical space occupied by the array of wires interferes with the beam. Calculations in SIMION \cite{Services2013} have shown that more than 98\% of particles are transmitted through the mesh detector without disturbance.

When the mesh detector is biased to ground, the detector acts as an anode inside a Faraday cup configuration. Upon ejection from CPET, 40-50\% of the electrons will hit the anode immediately so the detector provides confirmation that electrons are trapped in CPET.

Upon installation, a bunch of $2.5 \times 10^8$ electrons in CPET generated a signal of 16 $\mu$Vs and could be observed. Without any further tuning or optimization, this trapped bunch of electron plasma conforms to the best requirements laid out by Ke \emph{et al}, $\frac{N_{i}}{N_{e}} \leq 10^{-5}$, where $N_{e}$ is the number of electrons and $N_i$ is the number of ions whose value should not be greater than $10^3$ \cite{Ke2007}. Later, with some tuning, signals of greater than $10^9$ electrons were observed. The next step for CPET includes the simultaneous trapping of electrons and HCIs.

%

\section{Conclusion and Outlook}
\label{sec:outlook}
TITAN is well-positioned to take advantage of the advances in the EBIT and CPET detailed in Sections \ref{sec:EBIT} \& \ref{sec:CPET}. Furthermore, a robust campaign of mass measurements of astrophysically important \emph{r}-process nuclei is ongoing.

A new Multi-Reflection Time-of-Flight mass spectrometer (MR-ToF) built at the Justus-Liebig-University in Gie{\ss}en, Germany was delivered to TRIUMF in October 2014. The Gie{\ss}en group demonstrated an offline mass resolution of $m/\Delta m = 50,000$ and a mass selective re-trapping resolution of $m/\Delta m = 13,000$ \cite{Jesch2015}. We are commissioning the MR-ToF locally before installing the device into the TITAN system.

In-trap decay spectroscopy will continue with several approved experiments, including a study of the change in the electron-capture decay lifetime for He- and H-like $^{64}$Cu. With the suppression of 511 keV $\beta$'s, new configurations for the detector array surrounding EBIT are being explored, including the substitution of some Si(Li) detectors for HPGe detectors which would provide increased detection efficiencies for higher-energy photons \cite{Leach2015b}.

CPET's new mesh detector is a polarity-agnostic detector solution for experimental regions with tight space constraints and/or high magnetic field environments. Its transparent nature alleviates the need for remote detector manipulation which saves on both cost and complexity, and its small profile makes its installation practical in almost any experimental setting. Photochemical etching is now a precise enough tool that custom anode shapes can be quickly fabricated for minimal cost.

With the successful installation of an in-line, effectively transparent detector that will confirm the presence of trapped electrons, CPET can begin the task of simultaneously trapping ions. A stable Cs surface source is ready for installation into the CPET beamline, and the electron source will be moved off the trap axis to allow for the passage of ions. Once those tasks are complete, CPET can be used for the simultaneous trapping of electrons and positively charged ions followed by the characterization of their interaction and a demonstration of cooling. Upon the successful demonstration of cooling, CPET will be installed into the TITAN beamline.


\section{Acknowledgements}
TRIUMF receives federal funding via a contribution agreement with the National Research Council of Canada (NRC). This work was partially supported by the Natural Sciences and Engineering Research Council of Canada (NSERC), the Canada Foundation for Innovation (CFI) and the Deutsche Forschungsgemeinschaft (DFG) under Grant FR 601/3-1. Author J. Even gratefully acknowledges the financial support from the German Academic Exchange Service (DAAD Postdoc program).

\section*{References}
\bibliography{library}

\begin{thebibliography}{44}
\expandafter\ifx\csname natexlab\endcsname\relax\def\natexlab#1{#1}\fi
\providecommand{\url}[1]{\texttt{#1}}
\providecommand{\href}[2]{#2}
\providecommand{\path}[1]{#1}
\providecommand{\DOIprefix}{doi:}
\providecommand{\ArXivprefix}{arXiv:}
\providecommand{\URLprefix}{URL: }
\providecommand{\Pubmedprefix}{pmid:}
\providecommand{\doi}[1]{\href{http://dx.doi.org/#1}{\path{#1}}}
\providecommand{\Pubmed}[1]{\href{pmid:#1}{\path{#1}}}
\providecommand{\bibinfo}[2]{#2}
\ifx\xfnm\relax \def\xfnm[#1]{\unskip,\space#1}\fi
\bibitem[{Schatz(2013)}]{Schatz2013}
\bibinfo{author}{H.~Schatz}, \bibinfo{journal}{Int. J. Mass Spectrom.}
  \bibinfo{volume}{349-350} (\bibinfo{year}{2013}) \bibinfo{pages}{181--186}.
  \DOIprefix\doi{10.1016/j.ijms.2013.03.016}.
\bibitem[{Lunney and Thibault(2003)}]{Lunney2003}
\bibinfo{author}{D.~Lunney}, \bibinfo{author}{C.~Thibault},
  \bibinfo{journal}{Rev. Mod. Phys.} \bibinfo{volume}{75}
  (\bibinfo{year}{2003}) \bibinfo{pages}{1021--1082}.
  \DOIprefix\doi{10.1103/RevModPhys.75.1021}.
\bibitem[{Gabrielse et~al.(1990)Gabrielse, Fei, Orozco, Tjoelker, Haas,
  Kalinowsky, Trainor, and Kells}]{Diffusion1990}
\bibinfo{author}{G.~Gabrielse}, \bibinfo{author}{X.~Fei},
  \bibinfo{author}{L.~Orozco}, \bibinfo{author}{R.~Tjoelker},
  \bibinfo{author}{J.~Haas}, \bibinfo{author}{H.~Kalinowsky},
  \bibinfo{author}{T.~Trainor}, \bibinfo{author}{W.~Kells},
  \bibinfo{journal}{Phys. Rev. Lett.} \bibinfo{volume}{65}
  (\bibinfo{year}{1990}) \bibinfo{pages}{1317--1320}.
  \DOIprefix\doi{10.1103/PhysRevLett.65.1317}.
\bibitem[{Blaum et~al.(2013)Blaum, Dilling, and
  N{\"{o}}rtersh{\"{a}}user}]{Blaum2012}
\bibinfo{author}{K.~Blaum}, \bibinfo{author}{J.~Dilling},
  \bibinfo{author}{W.~N{\"{o}}rtersh{\"{a}}user}, \bibinfo{journal}{Phys. Scr.}
  \bibinfo{volume}{T152} (\bibinfo{year}{2013}) \bibinfo{pages}{014017}.
  \DOIprefix\doi{10.1088/0031-8949/2013/T152/014017}.
\bibitem[{Valverde et~al.(2015)Valverde, Bollen, Brodeur, Bryce, Cooper,
  Eibach, Gulyuz, Izzo, Morrissey, Redshaw, Ringle, Sandler, Schwarz,
  Sumithrarachchi, and Villari}]{Valverde2015}
\bibinfo{author}{A.~A. Valverde}, \bibinfo{author}{G.~Bollen},
  \bibinfo{author}{M.~Brodeur}, \bibinfo{author}{R.~A. Bryce},
  \bibinfo{author}{K.~Cooper}, \bibinfo{author}{M.~Eibach},
  \bibinfo{author}{K.~Gulyuz}, \bibinfo{author}{C.~Izzo},
  \bibinfo{author}{D.~J. Morrissey}, \bibinfo{author}{M.~Redshaw},
  \bibinfo{author}{R.~Ringle}, \bibinfo{author}{R.~Sandler},
  \bibinfo{author}{S.~Schwarz}, \bibinfo{author}{C.~S. Sumithrarachchi},
  \bibinfo{author}{A.~C.~C. Villari}, \bibinfo{journal}{Phys. Rev. Lett.}
  \bibinfo{volume}{114} (\bibinfo{year}{2015}) \bibinfo{pages}{232502}.
  \DOIprefix\doi{10.1103/PhysRevLett.114.232502}.
\bibitem[{{Van Schelt} et~al.(2013){Van Schelt}, Lascar, Savard, Clark,
  Bertone, Caldwell, Chaudhuri, Levand, Li, Morgan, Orford, Segel, Sharma, and
  Sternberg}]{VanSchelt2013}
\bibinfo{author}{J.~{Van Schelt}}, \bibinfo{author}{D.~Lascar},
  \bibinfo{author}{G.~Savard}, \bibinfo{author}{J.~A. Clark},
  \bibinfo{author}{P.~F. Bertone}, \bibinfo{author}{S.~Caldwell},
  \bibinfo{author}{A.~Chaudhuri}, \bibinfo{author}{A.~F. Levand},
  \bibinfo{author}{G.~Li}, \bibinfo{author}{G.~E. Morgan},
  \bibinfo{author}{R.~Orford}, \bibinfo{author}{R.~E. Segel},
  \bibinfo{author}{K.~S. Sharma}, \bibinfo{author}{M.~G. Sternberg},
  \bibinfo{journal}{Phys. Rev. Lett.} \bibinfo{volume}{111}
  (\bibinfo{year}{2013}) \bibinfo{pages}{061102}.
  \DOIprefix\doi{10.1103/PhysRevLett.111.061102}.
\bibitem[{Kankainen et~al.(2013)Kankainen, Hakala, Eronen, Gorelov, Jokinen,
  Kolhinen, Moore, Penttil{\"{a}}, Rinta-Antila, Rissanen, Saastamoinen,
  Sonnenschein, and {\"{A}}yst{\"{o}}}]{Kankainen2013}
\bibinfo{author}{A.~Kankainen}, \bibinfo{author}{J.~Hakala},
  \bibinfo{author}{T.~Eronen}, \bibinfo{author}{D.~Gorelov},
  \bibinfo{author}{A.~Jokinen}, \bibinfo{author}{V.~S. Kolhinen},
  \bibinfo{author}{I.~D. Moore}, \bibinfo{author}{H.~Penttil{\"{a}}},
  \bibinfo{author}{S.~Rinta-Antila}, \bibinfo{author}{J.~Rissanen},
  \bibinfo{author}{A.~Saastamoinen}, \bibinfo{author}{V.~Sonnenschein},
  \bibinfo{author}{J.~{\"{A}}yst{\"{o}}}, \bibinfo{journal}{Phys. Rev. C}
  \bibinfo{volume}{87} (\bibinfo{year}{2013}) \bibinfo{pages}{024307}.
  \DOIprefix\doi{10.1103/PhysRevC.87.024307}.
\bibitem[{Macdonald et~al.(2014)Macdonald, Schultz, Bale, Chaudhuri, Chowdhury,
  Frekers, Gallant, Grossheim, Kwiatkowski, Lennarz, Simon, Simon, and
  Dilling}]{Macdonald2014}
\bibinfo{author}{T.~D. Macdonald}, \bibinfo{author}{B.~E. Schultz},
  \bibinfo{author}{J.~C. Bale}, \bibinfo{author}{A.~Chaudhuri},
  \bibinfo{author}{U.~Chowdhury}, \bibinfo{author}{D.~Frekers},
  \bibinfo{author}{A.~T. Gallant}, \bibinfo{author}{A.~Grossheim},
  \bibinfo{author}{A.~A. Kwiatkowski}, \bibinfo{author}{A.~Lennarz},
  \bibinfo{author}{M.~C. Simon}, \bibinfo{author}{V.~V. Simon},
  \bibinfo{author}{J.~Dilling}, \bibinfo{journal}{Phys. Rev. C}
  \bibinfo{volume}{89} (\bibinfo{year}{2014}) \bibinfo{pages}{044318}.
  \DOIprefix\doi{10.1103/PhysRevC.89.044318}.
\bibitem[{Droese et~al.(2013)Droese, Ackermann, Andersson, Blaum, Block,
  Dworschak, Eibach, Eliseev, Forsberg, Haettner, Herfurth, He{\ss}berger,
  Hofmann, Ketelaer, Marx, {Minaya Ramirez}, Nesterenko, Novikov, Pla{\ss},
  Rodr{\'{\i}}guez, Rudolph, Scheidenberger, Schweikhard, Stolze, Thirolf, and
  Weber}]{Droese2013}
\bibinfo{author}{C.~Droese}, \bibinfo{author}{D.~Ackermann},
  \bibinfo{author}{L.~L. Andersson}, \bibinfo{author}{K.~Blaum},
  \bibinfo{author}{M.~Block}, \bibinfo{author}{M.~Dworschak},
  \bibinfo{author}{M.~Eibach}, \bibinfo{author}{S.~Eliseev},
  \bibinfo{author}{U.~Forsberg}, \bibinfo{author}{E.~Haettner},
  \bibinfo{author}{F.~Herfurth}, \bibinfo{author}{F.~P. He{\ss}berger},
  \bibinfo{author}{S.~Hofmann}, \bibinfo{author}{J.~Ketelaer},
  \bibinfo{author}{G.~Marx}, \bibinfo{author}{E.~{Minaya Ramirez}},
  \bibinfo{author}{D.~Nesterenko}, \bibinfo{author}{Y.~N. Novikov},
  \bibinfo{author}{W.~R. Pla{\ss}}, \bibinfo{author}{D.~Rodr{\'{\i}}guez},
  \bibinfo{author}{D.~Rudolph}, \bibinfo{author}{C.~Scheidenberger},
  \bibinfo{author}{L.~Schweikhard}, \bibinfo{author}{S.~Stolze},
  \bibinfo{author}{P.~G. Thirolf}, \bibinfo{author}{C.~Weber},
  \bibinfo{journal}{Eur. Phys. J. A} \bibinfo{volume}{49}
  (\bibinfo{year}{2013}) \bibinfo{pages}{13}.
  \DOIprefix\doi{10.1140/epja/i2013-13013-0}.
\bibitem[{Dilling et~al.(2003)Dilling, Bricault, Smith, and
  Kluge}]{Dilling2003}
\bibinfo{author}{J.~Dilling}, \bibinfo{author}{P.~Bricault},
  \bibinfo{author}{M.~Smith}, \bibinfo{author}{H.~J. Kluge}, in:
  \bibinfo{booktitle}{Nucl. Instruments Methods Phys. Res. Sect. B Beam
  Interact. with Mater. Atoms}, volume \bibinfo{volume}{204}, pp.
  \bibinfo{pages}{492--496}. \DOIprefix\doi{10.1016/S0168-583X(02)02118-3}.
\bibitem[{Kwiatkowski et~al.(2013)Kwiatkowski, Macdonald, Andreoiu, Bale,
  Brunner, Chaudhuri, Chowdhury, Ettenauer, Gallant, Grossheim, Lennarz,
  Man{\'{e}}, Pearson, Schultz, Simon, Simon, and Dilling}]{Kwiatkowski2013}
\bibinfo{author}{A.~A. Kwiatkowski}, \bibinfo{author}{T.~D. Macdonald},
  \bibinfo{author}{C.~Andreoiu}, \bibinfo{author}{J.~C. Bale},
  \bibinfo{author}{T.~Brunner}, \bibinfo{author}{A.~Chaudhuri},
  \bibinfo{author}{U.~Chowdhury}, \bibinfo{author}{S.~Ettenauer},
  \bibinfo{author}{A.~Gallant}, \bibinfo{author}{A.~Grossheim},
  \bibinfo{author}{A.~Lennarz}, \bibinfo{author}{E.~Man{\'{e}}},
  \bibinfo{author}{M.~R. Pearson}, \bibinfo{author}{B.~E. Schultz},
  \bibinfo{author}{M.~C. Simon}, \bibinfo{author}{V.~V. Simon},
  \bibinfo{author}{J.~Dilling}, \bibinfo{journal}{Nucl. Instruments Methods
  Phys. Res. Sect. B Beam Interact. with Mater. Atoms} \bibinfo{volume}{317}
  (\bibinfo{year}{2013}) \bibinfo{pages}{517--521}.
  \DOIprefix\doi{10.1016/j.nimb.2013.05.087}.
\bibitem[{Dilling et~al.(2014)Dilling, Kruecken, and Merminga}]{Dilling2014}
\bibinfo{author}{J.~Dilling}, \bibinfo{author}{R.~Kruecken},
  \bibinfo{author}{L.~Merminga}, \bibinfo{title}{{ISAC and ARIEL: The TRIUMF
  Radioactive Beam Facilities and the Scientific Program}},
  \bibinfo{edition}{1st ed} ed., \bibinfo{publisher}{Springer Netherlands},
  \bibinfo{address}{Dordrecht}, \bibinfo{year}{2014}.
  \DOIprefix\doi{10.1007/978-94-007-7963-1}.
\bibitem[{Brunner et~al.(2012{\natexlab{a}})Brunner, Smith, Brodeur, Ettenauer,
  Gallant, Simon, Chaudhuri, Lapierre, Man{\'{e}}, Ringle, Simon, Vaz, Delheij,
  Good, Pearson, and Dilling}]{Brunner2012a}
\bibinfo{author}{T.~Brunner}, \bibinfo{author}{M.~Smith},
  \bibinfo{author}{M.~Brodeur}, \bibinfo{author}{S.~Ettenauer},
  \bibinfo{author}{A.~Gallant}, \bibinfo{author}{V.~Simon},
  \bibinfo{author}{A.~Chaudhuri}, \bibinfo{author}{A.~Lapierre},
  \bibinfo{author}{E.~Man{\'{e}}}, \bibinfo{author}{R.~Ringle},
  \bibinfo{author}{M.~Simon}, \bibinfo{author}{J.~Vaz},
  \bibinfo{author}{P.~Delheij}, \bibinfo{author}{M.~Good},
  \bibinfo{author}{M.~Pearson}, \bibinfo{author}{J.~Dilling},
  \bibinfo{journal}{Nucl. Instruments Methods Phys. Res. Sect. A Accel.
  Spectrometers, Detect. Assoc. Equip.} \bibinfo{volume}{676}
  (\bibinfo{year}{2012}{\natexlab{a}}) \bibinfo{pages}{32--43}.
  \DOIprefix\doi{10.1016/j.nima.2012.02.004}.
\bibitem[{Brunner et~al.(2012{\natexlab{b}})Brunner, Mueller, O’Sullivan,
  Simon, Kossick, Ettenauer, Gallant, Man{\'{e}}, Bishop, Good, Gratta, and
  Dilling}]{Brunner2012}
\bibinfo{author}{T.~Brunner}, \bibinfo{author}{A.~Mueller},
  \bibinfo{author}{K.~O’Sullivan}, \bibinfo{author}{M.~Simon},
  \bibinfo{author}{M.~Kossick}, \bibinfo{author}{S.~Ettenauer},
  \bibinfo{author}{A.~Gallant}, \bibinfo{author}{E.~Man{\'{e}}},
  \bibinfo{author}{D.~Bishop}, \bibinfo{author}{M.~Good},
  \bibinfo{author}{G.~Gratta}, \bibinfo{author}{J.~Dilling},
  \bibinfo{journal}{Int. J. Mass Spectrom.} \bibinfo{volume}{309}
  (\bibinfo{year}{2012}{\natexlab{b}}) \bibinfo{pages}{97--103}.
  \DOIprefix\doi{10.1016/j.ijms.2011.09.004}.
\bibitem[{Brodeur et~al.(2009)Brodeur, Brunner, Champagne, Ettenauer, Smith,
  Lapierre, Ringle, Ryjkov, Audi, Delheij, Lunney, and Dilling}]{Brodeur2009}
\bibinfo{author}{M.~Brodeur}, \bibinfo{author}{T.~Brunner},
  \bibinfo{author}{C.~Champagne}, \bibinfo{author}{S.~Ettenauer},
  \bibinfo{author}{M.~Smith}, \bibinfo{author}{A.~Lapierre},
  \bibinfo{author}{R.~Ringle}, \bibinfo{author}{V.~L. Ryjkov},
  \bibinfo{author}{G.~Audi}, \bibinfo{author}{P.~Delheij},
  \bibinfo{author}{D.~Lunney}, \bibinfo{author}{J.~Dilling},
  \bibinfo{journal}{Phys. Rev. C} \bibinfo{volume}{80} (\bibinfo{year}{2009})
  \bibinfo{pages}{044318}. \DOIprefix\doi{10.1103/PhysRevC.80.044318}.
\bibitem[{Brodeur et~al.(2012)Brodeur, Ryjkov, Brunner, Ettenauer, Gallant,
  Simon, Smith, Lapierre, Ringle, Delheij, Good, Lunney, and
  Dilling}]{Brodeur2012}
\bibinfo{author}{M.~Brodeur}, \bibinfo{author}{V.~Ryjkov},
  \bibinfo{author}{T.~Brunner}, \bibinfo{author}{S.~Ettenauer},
  \bibinfo{author}{A.~T. Gallant}, \bibinfo{author}{V.~Simon},
  \bibinfo{author}{M.~Smith}, \bibinfo{author}{A.~Lapierre},
  \bibinfo{author}{R.~Ringle}, \bibinfo{author}{P.~Delheij},
  \bibinfo{author}{M.~Good}, \bibinfo{author}{D.~Lunney},
  \bibinfo{author}{J.~Dilling}, \bibinfo{journal}{Int. J. Mass Spectrom.}
  \bibinfo{volume}{310} (\bibinfo{year}{2012}) \bibinfo{pages}{20--31}.
  \DOIprefix\doi{10.1016/j.ijms.2011.11.002}.
\bibitem[{Ringle et~al.(2007)Ringle, Bollen, Prinke, Savory, Schury, Schwarz,
  and Sun}]{Ringle2007}
\bibinfo{author}{R.~Ringle}, \bibinfo{author}{G.~Bollen},
  \bibinfo{author}{A.~Prinke}, \bibinfo{author}{J.~Savory},
  \bibinfo{author}{P.~Schury}, \bibinfo{author}{S.~Schwarz},
  \bibinfo{author}{T.~Sun}, \bibinfo{journal}{Int. J. Mass Spectrom.}
  \bibinfo{volume}{263} (\bibinfo{year}{2007}) \bibinfo{pages}{38--44}.
  \DOIprefix\doi{10.1016/j.ijms.2006.12.008}.
\bibitem[{Chaudhuri et~al.(2013)Chaudhuri, Andreoiu, Brunner, Chowdhury,
  Ettenauer, Gallant, Gwinner, Kwiatkowski, Lennarz, Lunney, Macdonald,
  Schultz, Simon, Simon, and Dilling}]{Chaudhuri2013}
\bibinfo{author}{A.~Chaudhuri}, \bibinfo{author}{C.~Andreoiu},
  \bibinfo{author}{T.~Brunner}, \bibinfo{author}{U.~Chowdhury},
  \bibinfo{author}{S.~Ettenauer}, \bibinfo{author}{A.~T. Gallant},
  \bibinfo{author}{G.~Gwinner}, \bibinfo{author}{A.~A. Kwiatkowski},
  \bibinfo{author}{A.~Lennarz}, \bibinfo{author}{D.~Lunney},
  \bibinfo{author}{T.~D. Macdonald}, \bibinfo{author}{B.~E. Schultz},
  \bibinfo{author}{M.~C. Simon}, \bibinfo{author}{V.~V. Simon},
  \bibinfo{author}{J.~Dilling}, \bibinfo{journal}{Phys. Rev. C - Nucl. Phys.}
  \bibinfo{volume}{88} (\bibinfo{year}{2013}) \bibinfo{pages}{1--5}.
  \DOIprefix\doi{10.1103/PhysRevC.88.054317}.
\bibitem[{Smith et~al.(2008)Smith, Brodeur, Brunner, Ettenauer, Lapierre,
  Ringle, Ryjkov, Ames, Bricault, Drake, Delheij, Lunney, Sarazin, and
  Dilling}]{Smith2008}
\bibinfo{author}{M.~Smith}, \bibinfo{author}{M.~Brodeur},
  \bibinfo{author}{T.~Brunner}, \bibinfo{author}{S.~Ettenauer},
  \bibinfo{author}{A.~Lapierre}, \bibinfo{author}{R.~Ringle},
  \bibinfo{author}{V.~L. Ryjkov}, \bibinfo{author}{F.~Ames},
  \bibinfo{author}{P.~Bricault}, \bibinfo{author}{G.~W.~F. Drake},
  \bibinfo{author}{P.~Delheij}, \bibinfo{author}{D.~Lunney},
  \bibinfo{author}{F.~Sarazin}, \bibinfo{author}{J.~Dilling},
  \bibinfo{journal}{Phys. Rev. Lett.} \bibinfo{volume}{101}
  (\bibinfo{year}{2008}) \bibinfo{pages}{202501}.
  \DOIprefix\doi{10.1103/PhysRevLett.101.202501}.
\bibitem[{Chaudhuri et~al.(2014)Chaudhuri, Andreoiu, Brodeur, Brunner,
  Chowdhury, Ettenauer, Gallant, Grossheim, Gwinner, Klawitter, Kwiatkowski,
  Leach, Lennarz, Lunney, Macdonald, Ringle, Schultz, Simon, Simon, and
  Dilling}]{Chaudhuri2014}
\bibinfo{author}{A.~Chaudhuri}, \bibinfo{author}{C.~Andreoiu},
  \bibinfo{author}{M.~Brodeur}, \bibinfo{author}{T.~Brunner},
  \bibinfo{author}{U.~Chowdhury}, \bibinfo{author}{S.~Ettenauer},
  \bibinfo{author}{A.~T. Gallant}, \bibinfo{author}{A.~Grossheim},
  \bibinfo{author}{G.~Gwinner}, \bibinfo{author}{R.~Klawitter},
  \bibinfo{author}{A.~A. Kwiatkowski}, \bibinfo{author}{K.~G. Leach},
  \bibinfo{author}{A.~Lennarz}, \bibinfo{author}{D.~Lunney},
  \bibinfo{author}{T.~D. Macdonald}, \bibinfo{author}{R.~Ringle},
  \bibinfo{author}{B.~E. Schultz}, \bibinfo{author}{V.~V. Simon},
  \bibinfo{author}{M.~C. Simon}, \bibinfo{author}{J.~Dilling},
  \bibinfo{journal}{Appl. Phys. B} \bibinfo{volume}{114} (\bibinfo{year}{2014})
  \bibinfo{pages}{99--105}. \DOIprefix\doi{10.1007/s00340-013-5618-8}.
\bibitem[{Lapierre et~al.(2010)Lapierre, Brodeur, Brunner, Ettenauer, Gallant,
  Simon, Good, Froese, {Crespo L{\'{o}}pez-Urrutia}, Delheij, Epp, Ringle,
  Schwarz, Ullrich, and Dilling}]{Lapierre2010}
\bibinfo{author}{A.~Lapierre}, \bibinfo{author}{M.~Brodeur},
  \bibinfo{author}{T.~Brunner}, \bibinfo{author}{S.~Ettenauer},
  \bibinfo{author}{A.~Gallant}, \bibinfo{author}{V.~Simon},
  \bibinfo{author}{M.~Good}, \bibinfo{author}{M.~Froese},
  \bibinfo{author}{J.~{Crespo L{\'{o}}pez-Urrutia}},
  \bibinfo{author}{P.~Delheij}, \bibinfo{author}{S.~Epp},
  \bibinfo{author}{R.~Ringle}, \bibinfo{author}{S.~Schwarz},
  \bibinfo{author}{J.~Ullrich}, \bibinfo{author}{J.~Dilling},
  \bibinfo{journal}{Nucl. Instruments Methods Phys. Res. Sect. A Accel.
  Spectrometers, Detect. Assoc. Equip.} \bibinfo{volume}{624}
  (\bibinfo{year}{2010}) \bibinfo{pages}{54--64}.
  \DOIprefix\doi{10.1016/j.nima.2010.09.030}.
\bibitem[{Ettenauer et~al.(2011)Ettenauer, Simon, Gallant, Brunner, Chowdhury,
  Simon, Brodeur, Chaudhuri, Man{\'{e}}, Andreoiu, Audi, L{\'{o}}pez-Urrutia,
  Delheij, Gwinner, Lapierre, Lunney, Pearson, Ringle, Ullrich, and
  Dilling}]{Ettenauer2011}
\bibinfo{author}{S.~Ettenauer}, \bibinfo{author}{M.~C. Simon},
  \bibinfo{author}{A.~T. Gallant}, \bibinfo{author}{T.~Brunner},
  \bibinfo{author}{U.~Chowdhury}, \bibinfo{author}{V.~V. Simon},
  \bibinfo{author}{M.~Brodeur}, \bibinfo{author}{A.~Chaudhuri},
  \bibinfo{author}{E.~Man{\'{e}}}, \bibinfo{author}{C.~Andreoiu},
  \bibinfo{author}{G.~Audi}, \bibinfo{author}{J.~R.~C. L{\'{o}}pez-Urrutia},
  \bibinfo{author}{P.~Delheij}, \bibinfo{author}{G.~Gwinner},
  \bibinfo{author}{A.~Lapierre}, \bibinfo{author}{D.~Lunney},
  \bibinfo{author}{M.~R. Pearson}, \bibinfo{author}{R.~Ringle},
  \bibinfo{author}{J.~Ullrich}, \bibinfo{author}{J.~Dilling},
  \bibinfo{journal}{Phys. Rev. Lett.} \bibinfo{volume}{107}
  (\bibinfo{year}{2011}) \bibinfo{pages}{272501}.
  \DOIprefix\doi{10.1103/PhysRevLett.107.272501}.
\bibitem[{Frekers et~al.(2013)Frekers, Simon, Andreoiu, Bale, Brodeur, Brunner,
  Chaudhuri, Chowdhury, {Crespo L{\'{o}}pez-Urrutia}, Delheij, Ejiri,
  Ettenauer, Gallant, Gavrin, Grossheim, Harakeh, Jang, Kwiatkowski, Lassen,
  Lennarz, Luichtl, Ma, Macdonald, Man{\'{e}}, Robertson, Schultz, Simon,
  Teigelh{\"{o}}fer, and Dilling}]{Frekers2013}
\bibinfo{author}{D.~Frekers}, \bibinfo{author}{M.~C. Simon},
  \bibinfo{author}{C.~Andreoiu}, \bibinfo{author}{J.~C. Bale},
  \bibinfo{author}{M.~Brodeur}, \bibinfo{author}{T.~Brunner},
  \bibinfo{author}{A.~Chaudhuri}, \bibinfo{author}{U.~Chowdhury},
  \bibinfo{author}{J.~R. {Crespo L{\'{o}}pez-Urrutia}},
  \bibinfo{author}{P.~Delheij}, \bibinfo{author}{H.~Ejiri},
  \bibinfo{author}{S.~Ettenauer}, \bibinfo{author}{A.~T. Gallant},
  \bibinfo{author}{V.~Gavrin}, \bibinfo{author}{A.~Grossheim},
  \bibinfo{author}{M.~N. Harakeh}, \bibinfo{author}{F.~Jang},
  \bibinfo{author}{A.~A. Kwiatkowski}, \bibinfo{author}{J.~Lassen},
  \bibinfo{author}{A.~Lennarz}, \bibinfo{author}{M.~Luichtl},
  \bibinfo{author}{T.~Ma}, \bibinfo{author}{T.~D. Macdonald},
  \bibinfo{author}{E.~Man{\'{e}}}, \bibinfo{author}{D.~Robertson},
  \bibinfo{author}{B.~E. Schultz}, \bibinfo{author}{V.~V. Simon},
  \bibinfo{author}{A.~Teigelh{\"{o}}fer}, \bibinfo{author}{J.~Dilling},
  \bibinfo{journal}{Phys. Lett. Sect. B Nucl. Elem. Part. High-Energy Phys.}
  \bibinfo{volume}{722} (\bibinfo{year}{2013}) \bibinfo{pages}{233--237}.
  \DOIprefix\doi{10.1016/j.physletb.2013.04.019}.
\bibitem[{Lapierre et~al.(2012)Lapierre, Brodeur, Brunner, Ettenauer, Finlay,
  Gallant, Simon, Delheij, Lunney, Ringle, Savajols, and
  Dilling}]{Lapierre2012}
\bibinfo{author}{A.~Lapierre}, \bibinfo{author}{M.~Brodeur},
  \bibinfo{author}{T.~Brunner}, \bibinfo{author}{S.~Ettenauer},
  \bibinfo{author}{P.~Finlay}, \bibinfo{author}{A.~T. Gallant},
  \bibinfo{author}{V.~V. Simon}, \bibinfo{author}{P.~Delheij},
  \bibinfo{author}{D.~Lunney}, \bibinfo{author}{R.~Ringle},
  \bibinfo{author}{H.~Savajols}, \bibinfo{author}{J.~Dilling},
  \bibinfo{journal}{Phys. Rev. C - Nucl. Phys.} \bibinfo{volume}{85}
  (\bibinfo{year}{2012}) \bibinfo{pages}{1--6}.
  \DOIprefix\doi{10.1103/PhysRevC.85.024317}.
\bibitem[{Gallant et~al.(2012)Gallant, Brodeur, Brunner, Chowdhury, Ettenauer,
  Simon, Man{\'{e}}, Simon, Andreoiu, Delheij, Gwinner, Pearson, Ringle, and
  Dilling}]{Gallant2012a}
\bibinfo{author}{A.~T. Gallant}, \bibinfo{author}{M.~Brodeur},
  \bibinfo{author}{T.~Brunner}, \bibinfo{author}{U.~Chowdhury},
  \bibinfo{author}{S.~Ettenauer}, \bibinfo{author}{V.~V. Simon},
  \bibinfo{author}{E.~Man{\'{e}}}, \bibinfo{author}{M.~C. Simon},
  \bibinfo{author}{C.~Andreoiu}, \bibinfo{author}{P.~Delheij},
  \bibinfo{author}{G.~Gwinner}, \bibinfo{author}{M.~R. Pearson},
  \bibinfo{author}{R.~Ringle}, \bibinfo{author}{J.~Dilling},
  \bibinfo{journal}{Phys. Rev. C - Nucl. Phys.} \bibinfo{volume}{85}
  (\bibinfo{year}{2012}) \bibinfo{pages}{1--7}.
  \DOIprefix\doi{10.1103/PhysRevC.85.044311}.
\bibitem[{Simon et~al.(2012)Simon, Brunner, Chowdhury, Eberhardt, Ettenauer,
  Gallant, Man{\'{e}}, Simon, Delheij, Pearson, Audi, Gwinner, Lunney, Schatz,
  and Dilling}]{Simon2012}
\bibinfo{author}{V.~V. Simon}, \bibinfo{author}{T.~Brunner},
  \bibinfo{author}{U.~Chowdhury}, \bibinfo{author}{B.~Eberhardt},
  \bibinfo{author}{S.~Ettenauer}, \bibinfo{author}{A.~T. Gallant},
  \bibinfo{author}{E.~Man{\'{e}}}, \bibinfo{author}{M.~C. Simon},
  \bibinfo{author}{P.~Delheij}, \bibinfo{author}{M.~R. Pearson},
  \bibinfo{author}{G.~Audi}, \bibinfo{author}{G.~Gwinner},
  \bibinfo{author}{D.~Lunney}, \bibinfo{author}{H.~Schatz},
  \bibinfo{author}{J.~Dilling}, \bibinfo{journal}{Phys. Rev. C - Nucl. Phys.}
  \bibinfo{volume}{85} (\bibinfo{year}{2012}) \bibinfo{pages}{1--11}.
  \DOIprefix\doi{10.1103/PhysRevC.85.064308}.
\bibitem[{Leach et~al.(2015)Leach, Grossheim, Lennarz, Brunner, {Crespo
  L{\'{o}}pez-Urrutia}, Gallant, Good, Klawitter, Kwiatkowski, Ma, Macdonald,
  Seeraji, Simon, Andreoiu, Dilling, and Frekers}]{Leach2015}
\bibinfo{author}{K.~Leach}, \bibinfo{author}{A.~Grossheim},
  \bibinfo{author}{A.~Lennarz}, \bibinfo{author}{T.~Brunner},
  \bibinfo{author}{J.~{Crespo L{\'{o}}pez-Urrutia}},
  \bibinfo{author}{A.~Gallant}, \bibinfo{author}{M.~Good},
  \bibinfo{author}{R.~Klawitter}, \bibinfo{author}{A.~A. Kwiatkowski},
  \bibinfo{author}{T.~Ma}, \bibinfo{author}{T.~Macdonald},
  \bibinfo{author}{S.~Seeraji}, \bibinfo{author}{M.~Simon},
  \bibinfo{author}{C.~Andreoiu}, \bibinfo{author}{J.~Dilling},
  \bibinfo{author}{D.~Frekers}, \bibinfo{journal}{Nucl. Instruments Methods
  Phys. Res. Sect. A Accel. Spectrometers, Detect. Assoc. Equip.}
  \bibinfo{volume}{780} (\bibinfo{year}{2015}) \bibinfo{pages}{91--99}.
  \DOIprefix\doi{10.1016/j.nima.2014.12.118}.
\bibitem[{Lennarz et~al.(2014)Lennarz, Grossheim, Leach, Alanssari, Brunner,
  Chaudhuri, Chowdhury, {Crespo L{\'{o}}pez-Urrutia}, Gallant, Holl,
  Kwiatkowski, Lassen, Macdonald, Schultz, Seeraji, Simon, Andreoiu, Dilling,
  and Frekers}]{Lennarz2014}
\bibinfo{author}{A.~Lennarz}, \bibinfo{author}{A.~Grossheim},
  \bibinfo{author}{K.~G. Leach}, \bibinfo{author}{M.~Alanssari},
  \bibinfo{author}{T.~Brunner}, \bibinfo{author}{A.~Chaudhuri},
  \bibinfo{author}{U.~Chowdhury}, \bibinfo{author}{J.~R. {Crespo
  L{\'{o}}pez-Urrutia}}, \bibinfo{author}{A.~T. Gallant},
  \bibinfo{author}{M.~Holl}, \bibinfo{author}{A.~A. Kwiatkowski},
  \bibinfo{author}{J.~Lassen}, \bibinfo{author}{T.~D. Macdonald},
  \bibinfo{author}{B.~E. Schultz}, \bibinfo{author}{S.~Seeraji},
  \bibinfo{author}{M.~C. Simon}, \bibinfo{author}{C.~Andreoiu},
  \bibinfo{author}{J.~Dilling}, \bibinfo{author}{D.~Frekers},
  \bibinfo{journal}{Phys. Rev. Lett.} \bibinfo{volume}{113}
  (\bibinfo{year}{2014}) \bibinfo{pages}{082502}.
  \DOIprefix\doi{10.1103/PhysRevLett.113.082502}.
\bibitem[{Leach et~al.(2015)Leach, Lennarz, Grossheim, Klawitter, Brunner,
  Chaudhuri, Chowdhury, {Crespo L{\'{o}}pez-Urrutia}, Gallant, Kwiatkowski,
  Macdonald, Schultz, Seeraji, Andreoiu, Frekers, and Dilling}]{Leach2015a}
\bibinfo{author}{K.~G. Leach}, \bibinfo{author}{A.~Lennarz},
  \bibinfo{author}{A.~Grossheim}, \bibinfo{author}{R.~Klawitter},
  \bibinfo{author}{T.~Brunner}, \bibinfo{author}{A.~Chaudhuri},
  \bibinfo{author}{U.~Chowdhury}, \bibinfo{author}{J.~R. {Crespo
  L{\'{o}}pez-Urrutia}}, \bibinfo{author}{A.~T. Gallant},
  \bibinfo{author}{A.~A. Kwiatkowski}, \bibinfo{author}{T.~D. Macdonald},
  \bibinfo{author}{B.~E. Schultz}, \bibinfo{author}{S.~Seeraji},
  \bibinfo{author}{C.~Andreoiu}, \bibinfo{author}{D.~Frekers},
  \bibinfo{author}{J.~Dilling}, in: \bibinfo{booktitle}{Proc. Conf. Adv.
  Radioact. Isot. Sci.}, volume \bibinfo{volume}{020040},
  \bibinfo{publisher}{Journal of the Physical Society of Japan},
  \bibinfo{year}{2015}, pp. \bibinfo{pages}{1--6}.
  \DOIprefix\doi{10.7566/JPSCP.6.020040}.
\bibitem[{Klawitter et~al.(2015)Klawitter, Alanssari, Chowdhury, Chaudhuri,
  {Crespo L{\'{o}}pez-Urrutia}, Ettenauer, Gallant, Grossheim, Gwinner,
  Kwiatkowski, Leach, Lennarz, Macdonald, Simon, Schultz, Seeraji, Andreoiu,
  Frekers, and Dilling}]{Klawitter2015a}
\bibinfo{author}{R.~Klawitter}, \bibinfo{author}{M.~Alanssari},
  \bibinfo{author}{U.~Chowdhury}, \bibinfo{author}{A.~Chaudhuri},
  \bibinfo{author}{J.~{Crespo L{\'{o}}pez-Urrutia}},
  \bibinfo{author}{S.~Ettenauer}, \bibinfo{author}{A.~T. Gallant},
  \bibinfo{author}{A.~Grossheim}, \bibinfo{author}{G.~Gwinner},
  \bibinfo{author}{A.~A. Kwiatkowski}, \bibinfo{author}{K.~Leach},
  \bibinfo{author}{A.~Lennarz}, \bibinfo{author}{T.~D. Macdonald},
  \bibinfo{author}{M.~C. Simon}, \bibinfo{author}{B.~E. Schultz},
  \bibinfo{author}{S.~Seeraji}, \bibinfo{author}{C.~Andreoiu},
  \bibinfo{author}{D.~Frekers}, \bibinfo{author}{J.~Dilling}, in:
  \bibinfo{booktitle}{AIP Conf. Proc.}, volume \bibinfo{volume}{1640}, pp.
  \bibinfo{pages}{112--119}. \DOIprefix\doi{10.1063/1.4905407}.
\bibitem[{Ke et~al.(2007)Ke, Shi, Gwinner, Sharma, Toews, Dilling, and
  Ryjkov}]{Ke2007}
\bibinfo{author}{Z.~Ke}, \bibinfo{author}{W.~Shi},
  \bibinfo{author}{G.~Gwinner}, \bibinfo{author}{K.~Sharma},
  \bibinfo{author}{S.~Toews}, \bibinfo{author}{J.~Dilling},
  \bibinfo{author}{V.~L. Ryjkov}, \bibinfo{journal}{Hyperfine Interact.}
  \bibinfo{volume}{173} (\bibinfo{year}{2007}) \bibinfo{pages}{103--111}.
  \DOIprefix\doi{10.1007/s10751-007-9548-x}.
\bibitem[{Froese(2006)}]{Froese2006a}
\bibinfo{author}{M.~W. Froese}, \bibinfo{title}{{The TITAN Electron Beam Ion
  Trap: Assembly, Characterization, and First Tests}}, \bibinfo{type}{Msc
  thesis}, University of Manitoba, \bibinfo{year}{2006}.
\bibitem[{Ettenauer et~al.(2013)Ettenauer, Simon, Macdonald, and
  Dilling}]{Ettenauer2013}
\bibinfo{author}{S.~Ettenauer}, \bibinfo{author}{M.~C. Simon},
  \bibinfo{author}{T.~D. Macdonald}, \bibinfo{author}{J.~Dilling},
  \bibinfo{journal}{Int. J. Mass Spectrom.} \bibinfo{volume}{349-350}
  (\bibinfo{year}{2013}) \bibinfo{pages}{74--80}.
  \DOIprefix\doi{10.1016/j.ijms.2013.04.021}.
\bibitem[{Schechter and Valle(1982)}]{Valle1982}
\bibinfo{author}{J.~Schechter}, \bibinfo{author}{J.~W.~F. Valle},
  \bibinfo{journal}{Phys. Rev. D} \bibinfo{volume}{25} (\bibinfo{year}{1982})
  \bibinfo{pages}{2951--2954}. \DOIprefix\doi{10.1103/PhysRevD.25.2951}.
\bibitem[{Takasugi(1984)}]{Takasugi1984}
\bibinfo{author}{E.~Takasugi}, \bibinfo{journal}{Phys. Lett. B}
  \bibinfo{volume}{149} (\bibinfo{year}{1984}) \bibinfo{pages}{372--376}.
  \DOIprefix\doi{10.1016/0370-2693(84)90426-X}.
\bibitem[{Barea et~al.(2012)Barea, Kotila, and Iachello}]{Barea2012}
\bibinfo{author}{J.~Barea}, \bibinfo{author}{J.~Kotila},
  \bibinfo{author}{F.~Iachello}, \bibinfo{journal}{Phys. Rev. Lett.}
  \bibinfo{volume}{109} (\bibinfo{year}{2012}) \bibinfo{pages}{042501}.
  \DOIprefix\doi{10.1103/PhysRevLett.109.042501}.
\bibitem[{Savard et~al.(1991)Savard, Becker, Bollen, Kluge, Moore, Otto,
  Schweikhard, Stolzenberg, and Wiess}]{Savard1991}
\bibinfo{author}{G.~Savard}, \bibinfo{author}{S.~Becker},
  \bibinfo{author}{G.~Bollen}, \bibinfo{author}{H.-J. Kluge},
  \bibinfo{author}{R.~Moore}, \bibinfo{author}{T.~Otto},
  \bibinfo{author}{L.~Schweikhard}, \bibinfo{author}{H.~Stolzenberg},
  \bibinfo{author}{U.~Wiess}, \bibinfo{journal}{Phys. Lett. A}
  \bibinfo{volume}{158} (\bibinfo{year}{1991}) \bibinfo{pages}{247--252}.
  \DOIprefix\doi{10.1016/0375-9601(91)91008-2}.
\bibitem[{Schultz et~al.(2013)Schultz, Chowdhury, Simon, Andreoiu, Chaudhuri,
  Gallant, Kwiatkowski, Macdonald, Simon, Dilling, and Gwinner}]{Schultz2013}
\bibinfo{author}{B.~E. Schultz}, \bibinfo{author}{U.~Chowdhury},
  \bibinfo{author}{V.~V. Simon}, \bibinfo{author}{C.~Andreoiu},
  \bibinfo{author}{A.~Chaudhuri}, \bibinfo{author}{A.~T. Gallant},
  \bibinfo{author}{A.~A. Kwiatkowski}, \bibinfo{author}{T.~D. Macdonald},
  \bibinfo{author}{M.~C. Simon}, \bibinfo{author}{J.~Dilling},
  \bibinfo{author}{G.~Gwinner}, \bibinfo{journal}{Phys. Scr.}
  \bibinfo{volume}{T156} (\bibinfo{year}{2013}) \bibinfo{pages}{014097}.
  \DOIprefix\doi{10.1088/0031-8949/2013/T156/014097}.
\bibitem[{Simon et~al.(2013)Simon, Macdonald, Bale, Chowdhury, Eberhardt,
  Eibach, Gallant, Jang, Lennarz, Luichtl, Ma, Robertson, Simon, Andreoiu,
  Brodeur, Brunner, Chaudhuri, {Crespo L{\'{o}}pez-Urrutia}, Delheij,
  Ettenauer, Frekers, Grossheim, Gwinner, Kwiatkowski, Lapierre, Man{\'{e}},
  Pearson, Ringle, Schultz, and Dilling}]{Simon2013}
\bibinfo{author}{M.~Simon}, \bibinfo{author}{T.~Macdonald},
  \bibinfo{author}{J.~C. Bale}, \bibinfo{author}{U.~Chowdhury},
  \bibinfo{author}{B.~Eberhardt}, \bibinfo{author}{M.~Eibach},
  \bibinfo{author}{A.~Gallant}, \bibinfo{author}{F.~Jang},
  \bibinfo{author}{A.~Lennarz}, \bibinfo{author}{M.~Luichtl},
  \bibinfo{author}{T.~Ma}, \bibinfo{author}{D.~Robertson},
  \bibinfo{author}{V.~V. Simon}, \bibinfo{author}{C.~Andreoiu},
  \bibinfo{author}{M.~Brodeur}, \bibinfo{author}{T.~Brunner},
  \bibinfo{author}{A.~Chaudhuri}, \bibinfo{author}{J.~R. {Crespo
  L{\'{o}}pez-Urrutia}}, \bibinfo{author}{P.~Delheij},
  \bibinfo{author}{S.~Ettenauer}, \bibinfo{author}{D.~Frekers},
  \bibinfo{author}{A.~Grossheim}, \bibinfo{author}{G.~Gwinner},
  \bibinfo{author}{A.~A. Kwiatkowski}, \bibinfo{author}{A.~Lapierre},
  \bibinfo{author}{E.~Man{\'{e}}}, \bibinfo{author}{M.~R. Pearson},
  \bibinfo{author}{R.~Ringle}, \bibinfo{author}{B.~E. Schultz},
  \bibinfo{author}{J.~Dilling}, \bibinfo{journal}{Phys. Scr.}
  \bibinfo{volume}{T156} (\bibinfo{year}{2013}) \bibinfo{pages}{014098}.
  \DOIprefix\doi{10.1088/0031-8949/2013/T156/014098}.
\bibitem[{Chowdhury et~al.(2015)Chowdhury, Good, Kootte, Lascar, Schultz,
  Dilling, and Gwinner}]{Chowdhury2015}
\bibinfo{author}{U.~Chowdhury}, \bibinfo{author}{M.~Good},
  \bibinfo{author}{B.~Kootte}, \bibinfo{author}{D.~Lascar},
  \bibinfo{author}{B.~E. Schultz}, \bibinfo{author}{J.~Dilling},
  \bibinfo{author}{G.~Gwinner}, in: \bibinfo{editor}{A.~Lapierre},
  \bibinfo{editor}{S.~Schwarz}, \bibinfo{editor}{T.~Baumann} (Eds.),
  \bibinfo{booktitle}{Proc. XII Int. Symp. ELECTRON BEAM ION SOURCES TRAPS},
  volume \bibinfo{volume}{1640}, \bibinfo{publisher}{AIP Press},
  \bibinfo{address}{East Lansing, Michigan}, \bibinfo{year}{2015}, pp.
  \bibinfo{pages}{120--123}. \DOIprefix\doi{10.1063/1.4905408}.
\bibitem[{Maero et~al.(2011)Maero, Paroli, Pozzoli, and Romé}]{Maero2011}
\bibinfo{author}{G.~Maero}, \bibinfo{author}{B.~Paroli},
  \bibinfo{author}{R.~Pozzoli}, \bibinfo{author}{M.~Romé},
  \bibinfo{journal}{Phys. Plasmas} \bibinfo{volume}{18} (\bibinfo{year}{2011})
  \bibinfo{pages}{032101}. \DOIprefix\doi{10.1063/1.3558374}.
\bibitem[{Scientific-Information-Services(2013)}]{Services2013}
\bibinfo{author}{Scientific-Information-Services}, \bibinfo{title}{{SIMION}},
  \bibinfo{year}{2013}.
\bibitem[{Jesch et~al.(2015)Jesch, Dickel, Pla{\ss}, Short, {Ayet San Andres},
  Dilling, Geissel, Greiner, Lang, Leach, Lippert, Scheidenberger, and
  Yavor}]{Jesch2015}
\bibinfo{author}{C.~Jesch}, \bibinfo{author}{T.~Dickel}, \bibinfo{author}{W.~R.
  Pla{\ss}}, \bibinfo{author}{D.~Short}, \bibinfo{author}{S.~{Ayet San
  Andres}}, \bibinfo{author}{J.~Dilling}, \bibinfo{author}{H.~Geissel},
  \bibinfo{author}{F.~Greiner}, \bibinfo{author}{J.~Lang},
  \bibinfo{author}{K.~G. Leach}, \bibinfo{author}{W.~Lippert},
  \bibinfo{author}{C.~Scheidenberger}, \bibinfo{author}{M.~I. Yavor}, in:
  \bibinfo{booktitle}{Proc. 6th Int. Conf. Trapped Charg. Part. Fundam. Phys.},
  \bibinfo{publisher}{Springer}, \bibinfo{address}{Takamatsu, Japan},
  \bibinfo{year}{2015}, pp. \bibinfo{pages}{1--10}.
  \DOIprefix\doi{10.1007/s10751-015-1184-2}.
\bibitem[{Leach et~al.(2015)Leach, Lennarz, Grossheim, Andreoiu, Dilling,
  Frekers, Good, and Seeraji}]{Leach2015b}
\bibinfo{author}{K.~G. Leach}, \bibinfo{author}{A.~Lennarz},
  \bibinfo{author}{A.~Grossheim}, \bibinfo{author}{C.~Andreoiu},
  \bibinfo{author}{J.~Dilling}, \bibinfo{author}{D.~Frekers},
  \bibinfo{author}{M.~Good}, \bibinfo{author}{S.~Seeraji},
  \bibinfo{journal}{EPJ Web Conf.} \bibinfo{volume}{93} (\bibinfo{year}{2015})
  \bibinfo{pages}{07006}. \DOIprefix\doi{10.1051/epjconf/20159307006}.

\end{thebibliography}

\end{document}